\def\unlock{\catcode`@=11} 
\def\lock{\catcode`@=12} 
\unlock
%
%
%
%
%

\font\fourteenrm=cmr10 scaled\magstep2
\font\twelverm=cmr10 scaled\magstep1
\font\ninerm=cmr9          \font\sixrm=cmr6

\font\fourteenbf=cmbx10 scaled\magstep2
\font\twelvebf=cmbx10 scaled\magstep1

\font\seventeeni=cmmi10 scaled\magstep3     \skewchar\seventeeni='177
\font\fourteeni=cmmi10 scaled\magstep2      \skewchar\fourteeni='177
\font\twelvei=cmmi10 scaled\magstep1        \skewchar\twelvei='177
\font\ninei=cmmi9                           \skewchar\ninei='177
\font\sixi=cmmi6                            \skewchar\sixi='177
\font\seventeensy=cmsy10 scaled\magstep3    \skewchar\seventeensy='60
\font\fourteensy=cmsy10 scaled\magstep2     \skewchar\fourteensy='60
\font\twelvesy=cmsy10 scaled\magstep1       \skewchar\twelvesy='60
\font\ninesy=cmsy9                          \skewchar\ninesy='60
\font\sixsy=cmsy6                           \skewchar\sixsy='60

\font\fourteenex=cmex10 scaled\magstep2
\font\twelveex=cmex10 scaled\magstep1

\font\fourteensl=cmsl10 scaled\magstep2
\font\twelvesl=cmsl10 scaled\magstep1

\font\fourteenit=cmti10 scaled\magstep2
\font\twelveit=cmti10 scaled\magstep1
\font\twelvett=cmtt10 scaled\magstep1
\font\twelvecp=cmcsc10 scaled\magstep1
\font\tencp=cmcsc10
\newfam\cpfam
%
%
\newcount\f@ntkey            \f@ntkey=0
\def\samef@nt{\relax \ifcase\f@ntkey \rm \or\oldstyle \or\or
         \or\it \or\sl \or\bf \or\tt \or\caps \fi }
\def\fourteenpoint{\relax
    \textfont0=\fourteenrm          \scriptfont0=\tenrm
    \scriptscriptfont0=\sevenrm
     \def\rm{\fam0 \fourteenrm \f@ntkey=0 }\relax
    \textfont1=\fourteeni           \scriptfont1=\teni
    \scriptscriptfont1=\seveni
     \def\oldstyle{\fam1 \fourteeni\f@ntkey=1 }\relax
    \textfont2=\fourteensy          \scriptfont2=\tensy
    \scriptscriptfont2=\sevensy
    \textfont3=\fourteenex     \scriptfont3=\fourteenex
    \scriptscriptfont3=\fourteenex
    \def\it{\fam\itfam \fourteenit\f@ntkey=4 }\textfont\itfam=\fourteenit
    \def\sl{\fam\slfam \fourteensl\f@ntkey=5 }\textfont\slfam=\fourteensl
    \scriptfont\slfam=\tensl
    \def\bf{\fam\bffam \fourteenbf\f@ntkey=6 }\textfont\bffam=\fourteenbf
    \scriptfont\bffam=\tenbf     \scriptscriptfont\bffam=\sevenbf
    \def\tt{\fam\ttfam \twelvett \f@ntkey=7 }\textfont\ttfam=\twelvett
    \h@big=11.9\p@{} \h@Big=16.1\p@{} \h@bigg=20.3\p@{} \h@Bigg=24.5\p@{}
    \def\caps{\fam\cpfam \twelvecp \f@ntkey=8 }\textfont\cpfam=\twelvecp
    \setbox\strutbox=\hbox{\vrule height 12pt depth 5pt width\z@}
    \samef@nt}
\def\twelvepoint{\relax
    \textfont0=\twelverm          \scriptfont0=\ninerm
    \scriptscriptfont0=\sixrm
     \def\rm{\fam0 \twelverm \f@ntkey=0 }\relax
    \textfont1=\twelvei           \scriptfont1=\ninei
    \scriptscriptfont1=\sixi
     \def\oldstyle{\fam1 \twelvei\f@ntkey=1 }\relax
    \textfont2=\twelvesy          \scriptfont2=\ninesy
    \scriptscriptfont2=\sixsy
    \textfont3=\twelveex          \scriptfont3=\twelveex
    \scriptscriptfont3=\twelveex
    \def\it{\fam\itfam \twelveit \f@ntkey=4 }\textfont\itfam=\twelveit
    \def\sl{\fam\slfam \twelvesl \f@ntkey=5 }\textfont\slfam=\twelvesl
    \scriptfont\slfam=\ninerm
    \def\bf{\fam\bffam \twelvebf \f@ntkey=6 }\textfont\bffam=\twelvebf
    \scriptfont\bffam=\ninerm     \scriptscriptfont\bffam=\sixrm
    \def\tt{\fam\ttfam \twelvett \f@ntkey=7 }\textfont\ttfam=\twelvett
    \h@big=10.2\p@{}
    \h@Big=13.8\p@{}
    \h@bigg=17.4\p@{}
    \h@Bigg=21.0\p@{}
    \def\caps{\fam\cpfam \twelvecp \f@ntkey=8 }\textfont\cpfam=\twelvecp
    \setbox\strutbox=\hbox{\vrule height 10pt depth 4pt width\z@}
    \samef@nt}
\def\tenpoint{\relax
    \textfont0=\tenrm          \scriptfont0=\sevenrm
    \scriptscriptfont0=\fiverm
    \def\rm{\fam0 \tenrm \f@ntkey=0 }\relax
    \textfont1=\teni           \scriptfont1=\seveni
    \scriptscriptfont1=\fivei
    \def\oldstyle{\fam1 \teni \f@ntkey=1 }\relax
    \textfont2=\tensy          \scriptfont2=\sevensy
    \scriptscriptfont2=\fivesy
    \textfont3=\tenex          \scriptfont3=\tenex
    \scriptscriptfont3=\tenex
    \def\it{\fam\itfam \tenit \f@ntkey=4 }\textfont\itfam=\tenit
    \def\sl{\fam\slfam \tensl \f@ntkey=5 }\textfont\slfam=\tensl
    \def\bf{\fam\bffam \tenbf \f@ntkey=6 }\textfont\bffam=\tenbf
    \scriptfont\bffam=\sevenbf     \scriptscriptfont\bffam=\fivebf
    \def\tt{\fam\ttfam \tentt \f@ntkey=7 }\textfont\ttfam=\tentt
    \def\caps{\fam\cpfam \tencp \f@ntkey=8 }\textfont\cpfam=\tencp
    \setbox\strutbox=\hbox{\vrule height 8.5pt depth 3.5pt width\z@}
    \samef@nt}
%
%
%
%
\newdimen\h@big  \h@big=8.5\p@
\newdimen\h@Big  \h@Big=11.5\p@
\newdimen\h@bigg  \h@bigg=14.5\p@
\newdimen\h@Bigg  \h@Bigg=17.5\p@
\def\big#1{{\hbox{$\left#1\vbox to\h@big{}\right.\n@space$}}}
\def\Big#1{{\hbox{$\left#1\vbox to\h@Big{}\right.\n@space$}}}
\def\bigg#1{{\hbox{$\left#1\vbox to\h@bigg{}\right.\n@space$}}}
\def\Bigg#1{{\hbox{$\left#1\vbox to\h@Bigg{}\right.\n@space$}}}
%
%
%
\normalbaselineskip = 20pt plus 0.2pt minus 0.1pt
\normallineskip = 1.5pt plus 0.1pt minus 0.1pt
\normallineskiplimit = 1.5pt
\newskip\normaldisplayskip
\normaldisplayskip = 18pt plus 4pt minus 8pt
\newskip\normaldispshortskip
\normaldispshortskip = 5pt plus 4pt
\newskip\normalparskip
\normalparskip = 6pt plus 2pt minus 1pt
\newskip\skipregister
\skipregister = 5pt plus 2pt minus 1.5pt
\newif\ifsingl@    \newif\ifdoubl@
\newif\iftwelv@    \twelv@true
\def\singlespace{\singl@true\doubl@false\spaces@t}
\def\doublespace{\singl@false\doubl@true\spaces@t}
\def\normalspace{\singl@false\doubl@false\spaces@t}
\def\Tenpoint{\tenpoint\twelv@false\spaces@t}
\def\Twelvepoint{\twelvepoint\twelv@true\spaces@t}
\def\spaces@t{\relax%
 \iftwelv@ \ifsingl@\subspaces@t3:4;\else\subspaces@t1:1;\fi%
 \else \ifsingl@\subspaces@t3:5;\else\subspaces@t4:5;\fi \fi%
 \ifdoubl@ \multiply\baselineskip by 5%
 \divide\baselineskip by 4 \fi \unskip}
\def\subspaces@t#1:#2;{%
      \baselineskip = \normalbaselineskip%
      \multiply\baselineskip by #1 \divide\baselineskip by #2%
      \lineskip = \normallineskip%
      \multiply\lineskip by #1 \divide\lineskip by #2%
      \lineskiplimit = \normallineskiplimit%
      \multiply\lineskiplimit by #1 \divide\lineskiplimit by #2%
      \parskip = \normalparskip%
      \multiply\parskip by #1 \divide\parskip by #2%
      \abovedisplayskip = \normaldisplayskip%
      \multiply\abovedisplayskip by #1 \divide\abovedisplayskip by #2%
      \belowdisplayskip = \abovedisplayskip%
      \abovedisplayshortskip = \normaldispshortskip%
      \multiply\abovedisplayshortskip by #1%
        \divide\abovedisplayshortskip by #2%
      \belowdisplayshortskip = \abovedisplayshortskip%
      \advance\belowdisplayshortskip by \belowdisplayskip%
      \divide\belowdisplayshortskip by 2%
      \smallskipamount = \skipregister%
      \multiply\smallskipamount by #1 \divide\smallskipamount by #2%
      \medskipamount = \smallskipamount \multiply\medskipamount by 2%
      \bigskipamount = \smallskipamount \multiply\bigskipamount by 4 }
\def\normalbaselines{ \baselineskip=\normalbaselineskip%
   \lineskip=\normallineskip \lineskiplimit=\normallineskip%
   \iftwelv@\else \multiply\baselineskip by 4 \divide\baselineskip by 5%
     \multiply\lineskiplimit by 4 \divide\lineskiplimit by 5%
     \multiply\lineskip by 4 \divide\lineskip by 5 \fi }
\Twelvepoint  
\interlinepenalty=50
\interfootnotelinepenalty=5000
\predisplaypenalty=9000
\postdisplaypenalty=500
\hfuzz=1pt
\vfuzz=0.2pt
%
%
%
\def\pagecontents{%
   \ifvoid\topins\else\unvbox\topins\vskip\skip\topins\fi
   \dimen@ = \dp255 \unvbox255
   \ifvoid\footins\else\vskip\skip\footins\footrule\unvbox\footins\fi
   \ifr@ggedbottom \kern-\dimen@ \vfil \fi }
\def\makeheadline{\vbox to 0pt{ \skip@=\topskip
      \advance\skip@ by -12pt \advance\skip@ by -2\normalbaselineskip
      \vskip\skip@ \line{\vbox to 12pt{}\the\headline} \vss
      }\nointerlineskip}
\def\makefootline{\baselineskip = 1.5\normalbaselineskip
                 \line{\the\footline}}
\newif\iffrontpage
\newif\ifletterstyle
\newif\ifp@genum
\def\nopagenumbers{\p@genumfalse}
\def\pagenumbers{\p@genumtrue}
\pagenumbers
\newtoks\paperheadline
\newtoks\letterheadline
\newtoks\letterfrontheadline
\newtoks\lettermainheadline
\newtoks\paperfootline
\newtoks\letterfootline
\newtoks\date
\footline={\ifletterstyle\the\letterfootline\else\the\paperfootline\fi}
\paperfootline={\hss\iffrontpage\else\ifp@genum\tenrm\folio\hss\fi\fi}
\letterfootline={\hfil}
\headline={\ifletterstyle\the\letterheadline\else\the\paperheadline\fi}
\paperheadline={\hfil}
\letterheadline{\iffrontpage\the\letterfrontheadline
     \else\the\lettermainheadline\fi}
\lettermainheadline={\rm\ifp@genum page \ \folio\fi\hfil\the\date}
\def\monthname{\relax\ifcase\month 0/\or January\or February\or
   March\or April\or May\or June\or July\or August\or September\or
   October\or November\or December\else\number\month/\fi}
\date={\monthname\ \number\day, \number\year}
\countdef\pagenumber=1  \pagenumber=1
\def\advancepageno{\global\advance\pageno by 1
   \ifnum\pagenumber<0 \global\advance\pagenumber by -1
    \else\global\advance\pagenumber by 1 \fi \global\frontpagefalse }
\def\folio{\ifnum\pagenumber<0 \romannumeral-\pagenumber
           \else \number\pagenumber \fi }
\def\footrule{\dimen@=\prevdepth\nointerlineskip
   \vbox to 0pt{\vskip -0.25\baselineskip \hrule width 0.35\hsize \vss}
   \prevdepth=\dimen@ }
\newtoks\foottokens
\foottokens={\Tenpoint\singlespace}
\newdimen\footindent
\footindent=24pt
\def\vfootnote#1{\insert\footins\bgroup  \the\foottokens
   \interlinepenalty=\interfootnotelinepenalty \floatingpenalty=20000
   \splittopskip=\ht\strutbox \boxmaxdepth=\dp\strutbox
   \leftskip=\footindent \rightskip=\z@skip
   \parindent=0.5\footindent \parfillskip=0pt plus 1fil
   \spaceskip=\z@skip \xspaceskip=\z@skip
   \Textindent{$ #1 $}\footstrut\futurelet\next\fo@t}
\def\Textindent#1{\noindent\llap{#1\enspace}\ignorespaces}
\def\footnote#1{\attach{#1}\vfootnote{#1}}

\let\footsymbol=\star
\newcount\lastf@@t           \lastf@@t=-1
\newcount\footsymbolcount    \footsymbolcount=0
\newif\ifPhysRev
\def\footsymbolgen{\relax \ifPhysRev \iffrontpage \NPsymbolgen\else
      \PRsymbolgen\fi \else \NPsymbolgen\fi
   \global\lastf@@t=\pageno \footsymbol }
\def\NPsymbolgen{\ifnum\footsymbolcount<0 \global\footsymbolcount=0\fi
   {\iffrontpage \else \advance\lastf@@t by 1 \fi
    \ifnum\lastf@@t<\pageno \global\footsymbolcount=0
     \else \global\advance\footsymbolcount by 1 \fi }
   \ifcase\footsymbolcount \fd@f\star\or \fd@f\dagger\or \fd@f\ddagger\or
    \fd@f\ast\or \fd@f\natural\or \fd@f\diamond\or \fd@f\bullet\or
    \fd@f\nabla\else \fd@f\dagger\global\footsymbolcount=0 \fi }
\def\fd@f#1{\xdef\footsymbol{#1}}
\def\PRsymbolgen{\ifnum\footsymbolcount>0 \global\footsymbolcount=0\fi
      \global\advance\footsymbolcount by -1
      \xdef\footsymbol{\sharp\number-\footsymbolcount} }
\def\space@ver#1{\let\@sf=\empty \ifmmode #1\else \ifhmode
   \edef\@sf{\spacefactor=\the\spacefactor}\unskip${}#1$\relax\fi\fi}
\def\attach#1{\space@ver{\strut^{\mkern 2mu #1} }\@sf\ }
%
%
%
\newcount\chapternumber      \chapternumber=0
\newcount\sectionnumber      \sectionnumber=0
\newcount\equanumber         \equanumber=0
\let\chapterlabel=0
\newtoks\chapterstyle        \chapterstyle={\Number}
\newskip\chapterskip         \chapterskip=\bigskipamount
\newskip\sectionskip         \sectionskip=\medskipamount
\newskip\headskip            \headskip=8pt plus 3pt minus 3pt
\newdimen\chapterminspace    \chapterminspace=15pc
\newdimen\sectionminspace    \sectionminspace=10pc
\newdimen\referenceminspace  \referenceminspace=25pc
\def\chapterreset{\global\advance\chapternumber by 1
   \ifnum\equanumber<0 \else\global\equanumber=0\fi
   \sectionnumber=0 \makel@bel}
\def\makel@bel{\xdef\chapterlabel{%
\the\chapterstyle{\the\chapternumber}.}}
\def\sectionlabel{\number\sectionnumber \quad }
\def\alphabetic#1{\count255='140 \advance\count255 by #1\char\count255}
\def\Alphabetic#1{\count255='100 \advance\count255 by #1\char\count255}
\def\Roman#1{\uppercase\expandafter{\romannumeral #1}}
\def\roman#1{\romannumeral #1}
\def\Number#1{\number #1}
\def\unnumberedchapters{\let\makel@bel=\relax \let\chapterlabel=\relax
\let\sectionlabel=\relax \equanumber=-1 }
\def\titlestyle#1{\par\begingroup \interlinepenalty=9999
     \leftskip=0.02\hsize plus 0.23\hsize minus 0.02\hsize
     \rightskip=\leftskip \parfillskip=0pt
     \hyphenpenalty=9000 \exhyphenpenalty=9000
     \tolerance=9999 \pretolerance=9000
     \spaceskip=0.333em \xspaceskip=0.5em
     \iftwelv@\twelvepoint\fourteenrm\else\twelvepoint\fi
   \noindent #1\par\endgroup }
\def\spacecheck#1{\dimen@=\pagegoal\advance\dimen@ by -\pagetotal
   \ifdim\dimen@<#1 \ifdim\dimen@>0pt \vfil\break \fi\fi}
\def\chapter#1{\par \penalty-300 \vskip\chapterskip
   \spacecheck\chapterminspace
   \chapterreset \titlestyle{\chapterlabel \ #1}
   \nobreak\vskip\headskip \penalty 30000
   \wlog{\string\chapter\ \chapterlabel} }

\def\section#1{\par \ifnum\the\lastpenalty=30000\else
   \penalty-200\vskip\sectionskip \spacecheck\sectionminspace\fi
   \wlog{\string\section\ \chapterlabel \the\sectionnumber}
   \global\advance\sectionnumber by 1  \noindent
   {\caps\enspace\chapterlabel \sectionlabel #1}\par
   \nobreak\vskip\headskip \penalty 30000 }
\def\subsection#1{\par
   \ifnum\the\lastpenalty=30000\else \penalty-100\smallskip \fi
   \noindent\undertext{#1}\enspace \vadjust{\penalty5000}}

\def\undertext#1{\vtop{\hbox{#1}\kern 1pt \hrule}}
\def\APPENDIX#1#2{\par\penalty-300\vskip\chapterskip
   \spacecheck\chapterminspace \chapterreset \xdef\chapterlabel{#1}
   \titlestyle{APPENDIX #2} \nobreak\vskip\headskip \penalty 30000
   \wlog{\string\Appendix\ \chapterlabel} }
\def\Appendix#1{\APPENDIX{#1}{#1}}
\def\appendix{\APPENDIX{A}{}}
%
%
%

\newif\ifdraftmode
\draftmodefalse

\def\eqname#1{\relax \ifnum\equanumber<0
     \xdef#1{{(\number-\equanumber)}}\global\advance\equanumber by -1
    \else \global\advance\equanumber by 1
      \xdef#1{{(\chapterlabel \number\equanumber)}} \fi}
\def\eqn#1{\eqno\eqname{#1}#1\ifdraftmode\rlap{\sevenrm\ \string#1}\fi}


%
\def\eqinsert#1{\noalign{\dimen@=\prevdepth \nointerlineskip
   \setbox0=\hbox to\displaywidth{\hfil #1}
   \vbox to 0pt{\vss\hbox{$\!\box0\!$}\kern-0.5\baselineskip}
   \prevdepth=\dimen@}}
%

%

%

%
%
\def\GENITEM#1;#2{\par \hangafter=0 \hangindent=#1
    \Textindent{$ #2 $}\ignorespaces}
\outer\def\newitem#1=#2;{\gdef#1{\GENITEM #2;}}
\newdimen\itemsize                \itemsize=30pt
\newitem\item=1\itemsize;
\newitem\sitem=1.75\itemsize;     
\newitem\ssitem=2.5\itemsize;     
\outer\def\newlist#1=#2&#3&#4;{\toks0={#2}\toks1={#3}%
   \count255=\escapechar \escapechar=-1
   \alloc@0\list\countdef\insc@unt\listcount     \listcount=0
   \edef#1{\par
      \countdef\listcount=\the\allocationnumber
      \advance\listcount by 1
      \hangafter=0 \hangindent=#4
      \Textindent{\the\toks0{\listcount}\the\toks1}}
   \expandafter\expandafter\expandafter
    \edef\c@t#1{begin}{\par
      \countdef\listcount=\the\allocationnumber \listcount=1
      \hangafter=0 \hangindent=#4
      \Textindent{\the\toks0{\listcount}\the\toks1}}
   \expandafter\expandafter\expandafter
    \edef\c@t#1{con}{\par \hangafter=0 \hangindent=#4 \noindent}
   \escapechar=\count255}
\def\c@t#1#2{\csname\string#1#2\endcsname}
\newlist\point=\Number&.&1.0\itemsize;
\newlist\subpoint=(\alphabetic&)&1.75\itemsize;
\newlist\subsubpoint=(\roman&)&2.5\itemsize;
\let\spoint=\subpoint

%
%
%
\newcount\referencecount     \referencecount=0
\newif\ifreferenceopen       \newwrite\referencewrite
\newtoks\rw@toks
\def\NPrefmark#1{\attach{\scriptscriptstyle [ #1 ] }}
\let\PRrefmark=\attach
\def\refmark#1{\relax\ifPhysRev\PRrefmark{#1}\else\NPrefmark{#1}\fi}
\def\refend{\refmark{\number\referencecount}}
\newcount\lastrefsbegincount \lastrefsbegincount=0
\def\refsend{\refmark{\count255=\referencecount
   \advance\count255 by-\lastrefsbegincount
   \ifcase\count255 \number\referencecount
   \or \number\lastrefsbegincount,\number\referencecount
   \else \number\lastrefsbegincount-\number\referencecount \fi}}
\def\refch@ck{\chardef\rw@write=\referencewrite
   \ifreferenceopen \else \referenceopentrue
   \immediate\openout\referencewrite=referenc.tex \fi}
%
{\catcode`\^^M=\active 
  \gdef\obeyendofline{\catcode`\^^M\active \let^^M\ }}%
%
{\catcode`\^^M=\active 
  \gdef\ignoreendofline{\catcode`\^^M=5}}
{\obeyendofline\gdef\rw@start#1{\def\t@st{#1} \ifx\t@st\blankend%
\endgroup \@sf \relax \else \ifx\t@st\bl@nkend \endgroup \@sf \relax%
\else \rw@begin#1
\backtotext
\fi \fi } }
{\obeyendofline\gdef\rw@begin#1
{\def\n@xt{#1}\rw@toks={#1}\relax%
\rw@next}}
\def\blankend{}
{\obeylines\gdef\bl@nkend{
}}
\newif\iffirstrefline  \firstreflinetrue
\def\rwr@teswitch{\ifx\n@xt\blankend \let\n@xt=\rw@begin %
 \else\iffirstrefline \global\firstreflinefalse%
\immediate\write\rw@write{\noexpand\obeyendofline \the\rw@toks}%
\let\n@xt=\rw@begin%
      \else\ifx\n@xt\rw@@d \def\n@xt{\immediate\write\rw@write{%
        \noexpand\ignoreendofline}\endgroup \@sf}%
             \else \immediate\write\rw@write{\the\rw@toks}%
             \let\n@xt=\rw@begin\fi\fi \fi}
\def\rw@next{\rwr@teswitch\n@xt}
\def\rw@@d{\backtotext} \let\rw@end=\relax
\let\backtotext=\relax

\newdimen\refindent     \refindent=30pt
\def\refitem#1{\par \hangafter=0 \hangindent=\refindent \Textindent{#1}}
\def\REFNUM#1{\space@ver{}\refch@ck \firstreflinetrue%
 \global\advance\referencecount by 1 \xdef#1{\the\referencecount}}
\def\refnum#1{\space@ver{}\refch@ck \firstreflinetrue%
 \global\advance\referencecount by 1 \xdef#1{\the\referencecount}\refend}

\def\REF#1{\REFNUM#1%
 \immediate\write\referencewrite{%
            \noexpand\refitem{\ifdraftmode{\sevenrm 
               \noexpand\string\string#1\ }\fi#1.}}%
      \begingroup\obeyendofline\rw@start}
\def\ref{\refnum\?%
 \immediate\write\referencewrite{\noexpand\refitem{\?.}}%
\begingroup\obeyendofline\rw@start}
\def\Ref#1{\refnum#1%
 \immediate\write\referencewrite{\noexpand\refitem{#1.}}%
\begingroup\obeyendofline\rw@start}
\def\REFS#1{\REFNUM#1\global\lastrefsbegincount=\referencecount
\immediate\write\referencewrite{\noexpand\refitem{#1.}}%
\begingroup\obeyendofline\rw@start}
%

%
%

%
\newcount\figurecount     \figurecount=0
\newif\iffigureopen       \newwrite\figurewrite
\def\figch@ck{\chardef\rw@write=\figurewrite \iffigureopen\else
   \immediate\openout\figurewrite=figures.aux
   \figureopentrue\fi}
\def\FIGNUM#1{\space@ver{}\figch@ck \firstreflinetrue%
 \global\advance\figurecount by 1 \xdef#1{\the\figurecount}}
\def\FIG#1{\FIGNUM#1
   \immediate\write\figurewrite{\noexpand\refitem{#1.}}%
   \begingroup\obeyendofline\rw@start}
\def\fig{\FIGNUM\? fig.~\?%
\immediate\write\figurewrite{\noexpand\refitem{\?.}}%
\begingroup\obeyendofline\rw@start}
\def\figure{\FIGNUM\? figure~\?
   \immediate\write\figurewrite{\noexpand\refitem{\?.}}%
   \begingroup\obeyendofline\rw@start}
\def\Fig{\FIGNUM\? Fig.~\?%
\immediate\write\figurewrite{\noexpand\refitem{\?.}}%
\begingroup\obeyendofline\rw@start}
\def\Figure{\FIGNUM\? Figure~\?%
\immediate\write\figurewrite{\noexpand\refitem{\?.}}%
\begingroup\obeyendofline\rw@start}
\newcount\tablecount     \tablecount=0
\newif\iftableopen       \newwrite\tablewrite
\def\tabch@ck{\chardef\rw@write=\tablewrite \iftableopen\else
   \immediate\openout\tablewrite=tables.aux
   \tableopentrue\fi}
\def\TABNUM#1{\space@ver{}\tabch@ck \firstreflinetrue%
 \global\advance\tablecount by 1 \xdef#1{\the\tablecount}}
\def\TABLE#1{\TABNUM#1
   \immediate\write\tablewrite{\noexpand\refitem{#1.}}%
   \begingroup\obeyendofline\rw@start}
\def\Table{\TABNUM\? Table~\?%
\immediate\write\tablewrite{\noexpand\refitem{\?.}}%
\begingroup\obeyendofline\rw@start}
\newif\ifsymbolopen       \newwrite\symbolwrite
\def\symch@ck{\ifsymbolopen\else
   \immediate\openout\symbolwrite=symbols.aux
   \symbolopentrue\fi}
\def\symdef#1#2{\def#1{#2}%
      \symch@ck%
      \immediate\write\symbolwrite{$$ \hbox{\noexpand\string\string#1} 
                \noexpand\qquad\noexpand\longrightarrow\noexpand\qquad 
                           \string#1 $$}}
%
%

%
%
%
\def\masterreset{\global\pagenumber=1 \global\chapternumber=0
   \global\equanumber=0 \global\sectionnumber=0
   \global\referencecount=0 \global\figurecount=0 \global\tablecount=0 }
\def\FRONTPAGE{\ifvoid255\else\vfill\penalty-2000\fi
      \masterreset\global\frontpagetrue
      \global\lastf@@t=0 \global\footsymbolcount=0}

\def\paperstyle{\letterstylefalse\normalspace\papersize}
\def\letterstyle{\letterstyletrue\singlespace\lettersize}
\def\papersize{\hsize=35pc\vsize=50pc\hoffset=1pc\voffset=6pc
               \skip\footins=\bigskipamount}
\def\lettersize{\hsize=6.5in\vsize=8.5in\hoffset=0in\voffset=1in
   \skip\footins=\smallskipamount \multiply\skip\footins by 3 }
\paperstyle   
%
%
\def\MEMO{\letterstyle\FRONTPAGE \letterfrontheadline={\hfil}
    \line{\quad\fourteenrm NTC MEMORANDUM\hfil\twelverm\the\date\quad}
    \medskip \memod@f}

\def\memit@m#1{\smallskip \hangafter=0 \hangindent=1in
      \Textindent{\caps #1}}
\def\memod@f{\xdef\to{\memit@m{To:}}\xdef\from{\memit@m{From:}}%
     \xdef\topic{\memit@m{Topic:}}\xdef\subject{\memit@m{Subject:}}%
     \xdef\rule{\bigskip\hrule height 1pt\bigskip}}
\memod@f
\newskip\lettertopfil
\lettertopfil = 0pt plus 1.5in minus 0pt
\newskip\letterbottomfil
\letterbottomfil = 0pt plus 2.3in minus 0pt
\newskip\spskip \setbox0\hbox{\ } \spskip=-1\wd0
\def\addressee#1{\medskip\rightline{\the\date\hskip 30pt} \bigskip
   \vskip\lettertopfil
   \ialign to\hsize{\strut ##\hfil\tabskip 0pt plus \hsize \cr #1\crcr}
   \medskip\noindent\hskip\spskip}
\newskip\signatureskip       \signatureskip=40pt
\def\signed#1{\par \penalty 9000 \bigskip \dt@pfalse
  \everycr={\noalign{\ifdt@p\vskip\signatureskip\global\dt@pfalse\fi}}
  \setbox0=\vbox{\singlespace \halign{\tabskip 0pt \strut ##\hfil\cr
   \noalign{\global\dt@ptrue}#1\crcr}}
  \line{\hskip 0.5\hsize minus 0.5\hsize \box0\hfil} \medskip }

\def\endletter{\ifnum\pagenumber=1 \vskip\letterbottomfil\supereject
\else \vfil\supereject \fi}
\newbox\letterb@x
\def\lettertext{\par\unvcopy\letterb@x\par}
\def\multiletter{\setbox\letterb@x=\vbox\bgroup
      \everypar{\vrule height 1\baselineskip depth 0pt width 0pt }
      \singlespace \topskip=\baselineskip }
\def\letterend{\par\egroup}
%
%
%
\newskip\frontpageskip
\newtoks\pubtype
\newtoks\Pubnum
\newtoks\pubnum
\newif\ifp@bblock  \p@bblocktrue
\def\PH@SR@V{\doubl@true \baselineskip=24.1pt plus 0.2pt minus 0.1pt
             \parskip= 3pt plus 2pt minus 1pt }
\def\PHYSREV{\paperstyle\PhysRevtrue\PH@SR@V}
\def\titlepage{\FRONTPAGE\paperstyle\ifPhysRev\PH@SR@V\fi
   \ifp@bblock\p@bblock\fi}
\def\nopubblock{\p@bblockfalse}
\def\endpage{\vfil\break}
\frontpageskip=1\medskipamount plus .5fil
\pubtype={\tensl Preliminary Version}
\Pubnum={$\caps SLAC - PUB - \the\pubnum $}
\pubnum={0000}
\def\p@bblock{\begingroup \tabskip=\hsize minus \hsize
   \baselineskip=1.5\ht\strutbox \topspace-2\baselineskip
   \halign to\hsize{\strut ##\hfil\tabskip=0pt\crcr
   \the\Pubnum\cr \the\date\cr \the\pubtype\cr}\endgroup}
\def\title#1{\vskip\frontpageskip \titlestyle{#1} \vskip\headskip }
\def\author#1{\vskip\frontpageskip\titlestyle{\twelvecp #1}\nobreak}

\def\address#1{\par\kern 5pt\titlestyle{\twelvepoint\it #1}}
\def\andaddress{\par\kern 5pt \centerline{\sl and} \address}

\def\abstract{\vskip\frontpageskip\centerline{\fourteenrm ABSTRACT}
              \vskip\headskip }

%
%
%

\def\\{\relax\ifmmode\backslash\else$\backslash$\fi}
\def\globaleqnumbers{\relax\if\equanumber<0\else\global\equanumber=-1\fi}

\def\journal#1&#2(#3){\unskip, \sl #1~\bf #2 \rm (19#3) }

\def\topspace{\hrule height 0pt depth 0pt \vskip}

\let\int=\intop         
\def\prop{\mathrel{{\mathchoice{\pr@p\scriptstyle}{\pr@p\scriptstyle}{
                \pr@p\scriptscriptstyle}{\pr@p\scriptscriptstyle} }}}
\def\pr@p#1{\setbox0=\hbox{$\cal #1 \char'103$}
   \hbox{$\cal #1 \char'117$\kern-.4\wd0\box0}}
\def\lsim{\mathrel{\mathpalette\@versim<}}
\def\gsim{\mathrel{\mathpalette\@versim>}}
\def\@versim#1#2{\lower0.5ex\vbox{\baselineskip\z@skip\lineskip-.1ex
  \lineskiplimit\z@\ialign{$\m@th#1\hfil##\hfil$\crcr#2\crcr\sim\crcr}}}
%
%
%
\let\sec@nt=\sec
\def\sec{\relax\ifmmode\let\n@xt=\sec@nt\else\let\n@xt\section\fi\n@xt}
\def\obsolete#1{\message{Macro \string #1 is obsolete.}}
\def\firstsec#1{\obsolete\firstsec \section{#1}}
\def\firstsubsec#1{\obsolete\firstsubsec \subsection{#1}}
\def\thispage#1{\obsolete\thispage \global\pagenumber=#1\frontpagefalse}
\def\thischapter#1{\obsolete\thischapter \global\chapternumber=#1}
\def\nextequation#1{\obsolete\nextequation \global\equanumber=#1
   \ifnum\the\equanumber>0 \global\advance\equanumber by 1 \fi}
\def\BOXITEM{\afterassigment\B@XITEM\setbox0=}
\def\B@XITEM{\par\hangindent\wd0 \noindent\box0 }
%

%
%
%
%
%
\lock
\message{   }
%
%
%












\newbox\figbox
\newdimen\zero  \zero=0pt
\newdimen\figmove
\newdimen\figwidth
\newdimen\figheight
\newdimen\textwidth
\newtoks\figtoks
\newcount\figcounta
\newcount\figcountb
\newcount\figlines
\def\figreset{\global\figcounta=-1 \global\figcountb=-1
\global\figmove=\baselineskip
\global\figlines=1 \global\figtoks={ } }
\def\picture#1by#2:#3{\global\setbox\figbox=\vbox{\vskip #1
\hbox{\vbox{\hsize=#2 \noindent #3}}}
\global\setbox\figbox=\vbox{\kern 10pt
\hbox{\kern 10pt \box\figbox \kern 10pt }\kern 10pt}
\global\figwidth=1\wd\figbox
\global\figheight=1\ht\figbox
\global\textwidth=\hsize
\global\advance\textwidth by - \figwidth }
\def\figtoksappend{\edef\temp##1{\global\figtoks=%
{\the\figtoks ##1}}\temp}
\def\figparmsa#1{\loop \global\advance\figcounta by 1
\ifnum \figcounta < #1
\figtoksappend{ 0pt \the\hsize }
\global\advance\figlines by 1
\repeat }
\def\figparmsb#1{\loop \global\advance\figcountb by 1
\ifnum \figcountb < #1
\figtoksappend{ \the\figwidth \the\textwidth}
\global\advance\figlines by 1
\repeat }
\def\figtext#1:#2:#3{\figreset%
\figparmsa{#1}%
\figparmsb{#2}%
\multiply\figmove by #1%
\global\setbox\figbox=\vbox to 0pt{\vskip \figmove  \hbox{\box\figbox}
\vss }
\parshape=\the\figlines\the\figtoks\the\zero\the\hsize
\noindent
\rlap{\box\figbox} #3}
\def\Buildrel#1\under#2{\mathrel{\mathop{#2}\limits_{#1}}}
\def\llongrarrow{\hbox to 40pt{\rightarrowfill}}



\def\boxit#1{\vbox{\hrule\hbox{\vrule\kern3pt
\vbox{\kern3pt#1\kern3pt}\kern3pt\vrule}\hrule}}
\newdimen\str
\def\fboxit#1#2{\vbox{\hrule height #1 \hbox{\vrule width #1
\kern3pt \vbox{\kern3pt#2\kern3pt}\kern3pt \vrule width #1 }
\hrule height #1 }}
\def\tran#1#2{\transpoint \hfuzz 5pt \gdef\label{#1}
\vbox to \the\vsize{\hsize \the\hsize #2} \par \eject }
\newdimen\baseskip
\newdimen\lskip
\lskip=\lineskip
\newdimen\transize
\newdimen\tall
\def\transpoint{\gdef\rm{\fam0\eighteenrm}
\font\twentyfourit = cmti10 scaled \magstep5
\font\twentyfourrm = cmr10 scaled \magstep5
\font\twentyfourbf = cmbx10 scaled \magstep5
\font\twentyeightsy = cmsy10 scaled \magstep5
\font\eighteenrm = cmr10 scaled \magstep3
\font\eighteenb = cmbx10 scaled \magstep3
\font\eighteeni = cmmi10 scaled \magstep3
\font\eighteenit = cmti10 scaled \magstep3
\font\eighteensl = cmsl10 scaled \magstep3
\font\eighteensy = cmsy10 scaled \magstep3
\font\eighteencaps = cmr10 scaled \magstep3
\font\eighteenmathex = cmex10 scaled \magstep3
\font\fourteenrm=cmr10 scaled \magstep2
\font\fourteeni=cmmi10 scaled \magstep2
\font\fourteenit = cmti10 scaled \magstep2
\font\fourteensy=cmsy10 scaled \magstep2
\font\fourteenmathex = cmex10 scaled \magstep2
\parindent 20pt
\global\hsize = 7.0in
\global\vsize = 8.9in
\dimen\transize = \the\hsize
\dimen\tall = \the\vsize
\def\sy{\eighteensy }
\def\sl{\eighteens }
\def\bf{\eighteenb }
\def\it{\eighteenit }
\def\caps{\eighteencaps }
\textfont0=\eighteenrm \scriptfont0=\fourteenrm
\scriptscriptfont0=\twelverm
\textfont1=\eighteeni \scriptfont1=\fourteeni \scriptscriptfont1=\twelvei
\textfont2=\eighteensy \scriptfont2=\fourteensy
\scriptscriptfont2=\twelvesy
\textfont3=\eighteenmathex \scriptfont3=\eighteenmathex
\scriptscriptfont3=\eighteenmathex
\global\baselineskip 35pt
\global\lineskip 15pt
\global\parskip 5pt  plus 1pt minus 1pt
\global\abovedisplayskip  3pt plus 10pt minus 10pt
\global\belowdisplayskip 3pt plus 10pt minus 10pt
\def\rtitle##1{\centerline{\undertext{\twentyfourrm ##1}}}
\def\ititle##1{\centerline{\undertext{\twentyfourit ##1}}}
\def\ctitle##1{\centerline{\undertext{\caps ##1}}}
\def\vstrut{\hbox{\vrule width 0pt height .35in depth .15in }}
\def\cline##1{\centerline{\vstrut ##1}}
\output{\shipout\vbox{\vskip .5in
\pagecontents \vfill
\hbox to \the\hsize{\hfill{\tenbf \label} } }
\global\advance\count0 by 1 }
\rm }


%
%
%

%

%

%

%
{\obeyspaces\global\let =\ }
%
%
%
\widowpenalty 1000
\thickmuskip 4mu plus 4mu

\def\bspac{\noalign{\vskip 4pt}}

\unlock
%
\Pubnum={${\twelverm IU/NTC}\  \the\pubnum $}
\pubnum={0000}
\def\p@nnlock{\begingroup \tabskip=\hsize minus \hsize
   \baselineskip=1.5\ht\strutbox \topspace-2\baselineskip
   \noindent\strut\the\Pubnum \hfill \the\date   \endgroup}
\def\titlepage{\FRONTPAGE\paperstyle\p@nnlock}
\def\displaylines#1{\displ@y
  \halign{\hbox to\displaywidth{$\hfil\displaystyle##\hfil$}\crcr
    #1\crcr}}
\def\addressee#1{\null
   \bigskip\medskip\rightline{\the\date\hskip 30pt}
   \vskip\lettertopfil
   \ialign to\hsize{\strut ##\hfil\tabskip 0pt plus \hsize \cr #1\crcr}
   \medskip\vskip 3pt\noindent}
\def\tmsaddressee#1#2{
   \vskip\lettertopfil
  \setbox0=\vbox{\singlespace \halign{\tabskip 0pt \strut ##\hfil\cr
   \noalign{\global\dt@ptrue}#1\crcr}}
  \line{\hskip 0.7\hsize minus 0.7\hsize \box0\hfil}
   \bigskip
   \vskip .2in
   \ialign to\hsize{\strut ##\hfil\tabskip 0pt plus \hsize \cr #2\crcr}
   \medskip\vskip 3pt\noindent}
\def\makeheadline{\vbox to 0pt{ \skip@=\topskip
      \advance\skip@ by -12pt \advance\skip@ by -2\normalbaselineskip
      \vskip\skip@  \vss
      }\nointerlineskip}
\def\signed#1{\par \penalty 9000 \bigskip \vskip .06in\dt@pfalse
  \everycr={\noalign{\ifdt@p\vskip\signatureskip\global\dt@pfalse\fi}}
  \setbox0=\vbox{\singlespace \halign{\tabskip 0pt \strut ##\hfil\cr
   \noalign{\global\dt@ptrue}#1\crcr}}
  \line{\hskip 0.5\hsize minus 0.5\hsize \box0\hfil} \medskip }
\def\lettersize{\hsize=6.25in\vsize=8.5in\hoffset=0in\voffset=1in
   \skip\footins=\smallskipamount \multiply\skip\footins by 3 }
%
%
%
%
%
\outer\def\newnewlist#1=#2&#3&#4&#5;{\toks0={#2}\toks1={#3}%
   \dimen1=\hsize  \advance\dimen1 by -#4
   \dimen2=\hsize  \advance\dimen2 by -#5
   \count255=\escapechar \escapechar=-1
   \alloc@0\list\countdef\insc@unt\listcount     \listcount=0
   \edef#1{\par
      \countdef\listcount=\the\allocationnumber
      \advance\listcount by 1
      \parshape=2 #4 \dimen1 #5 \dimen2
      \Textindent{\the\toks0{\listcount}\the\toks1}}
   \expandafter\expandafter\expandafter
    \edef\c@t#1{begin}{\par
      \countdef\listcount=\the\allocationnumber \listcount=1
      \parshape=2 #4 \dimen1 #5 \dimen2
      \Textindent{\the\toks0{\listcount}\the\toks1}}
   \expandafter\expandafter\expandafter
    \edef\c@t#1{con}{\par \parshape=2 #4 \dimen1 #5 \dimen2 \noindent}
   \escapechar=\count255}
\def\c@t#1#2{\csname\string#1#2\endcsname}
%
%
%
%
%
%
%
\def\noparGENITEM#1;{\hangafter=0 \hangindent=#1
    \ignorespaces\noindent}
\outer\def\noparnewitem#1=#2;{\gdef#1{\noparGENITEM #2;}}
\noparnewitem\spoint=1.5\itemsize;
%
%
%
\def\MEMO{\letterstyle\FRONTPAGE \letterfrontheadline={\hfil}
      \hoffset=1in \voffset=1.21in
    \line{\hskip .8in  \special{overlay ntcmemo.dat}
          \quad\fourteenrm NTC MEMORANDUM\hfil\twelverm\the\date\quad}
    \medskip\medskip \memod@f}

\def\memit@m#1{\smallskip \hangafter=0 \hangindent=1in
      \Textindent{\caps #1}}
\def\memod@f{\xdef\to{\memit@m{To:}}\xdef\from{\memit@m{From:}}%
     \xdef\topic{\memit@m{Topic:}}\xdef\subject{\memit@m{Subject:}}%
     \xdef\rule{\bigskip\hrule height 1pt\bigskip}}
\memod@f
\lock
%

%
%
%
%
%
%
%

\let\ssize=\scriptstyle
\let\sssize\scriptscriptstyle
%
%
%
%
%
\def\tildesymbol{\mathchar"0218 }
\def\undervec#1{\setbox0\hbox{$\tildesymbol$}\setbox1\hbox{$#1$}#1%
                \dimen0=\wd0\advance\dimen0by\wd1\divide\dimen0by2
                 \kern-\dimen0\lower1.3ex\hbox{$\tildesymbol$}}
\def\sundervec#1{\setbox0\hbox{$\ssize\tildesymbol$}
                 \setbox1\hbox{$\ssize #1$}#1%
                \dimen0=\wd0\advance\dimen0by\wd1\divide\dimen0by2
                 \kern-\dimen0\lower.85ex\hbox{$\ssize\tildesymbol$}}
\def\ssundervec#1{\setbox0\hbox{$\sssize\tildesymbol$}
                 \setbox1\hbox{$\sssize #1$}#1%
                \dimen0=\wd0\advance\dimen0by\wd1\divide\dimen0by2
                 \kern-\dimen0\lower.6ex\hbox{$\sssize\tildesymbol$}}
%

%


%

%

%

%

%

%

%

%
%
%
%
%

%

%

%

%

%

%
%
%
%
%
%

%

%

%

%

%
%
%
%
%
\def\intback{\kern-.1em}

%
%
%
%
%

%
\def\sixj#1#2#3#4#5#6{\left\{\matrix{#1 & #2 & #3 \cr
                                      #4 & #5 & #6 \cr}\right\}}

\def\bspac{\noalign{\vskip 4pt}}

%
%
%
%
%

%
%
%
%
%

%
%
%
%
%

%

%
%

%

%

%

%

%

%

%

%

%
%
%
%
%

%
%
%
%

%
%
%
\def\papersize{\hsize=6.25in\vsize=8.60in\hoffset=0.0in\voffset=0.1in
                \skip\footins=\bigskipamount}
\paperstyle
\doublespace
\pretolerance=500
\tolerance=500
\widowpenalty 5000
\thinmuskip=3mu
\medmuskip=4mu plus 2mu minus 4mu
\thickmuskip=5mu plus 5mu
\def\NPrefmark#1{\attach{\ssize #1{}}}
%
%
%
%
\def\NPrefmark#1{\attach{\scriptstyle  #1 }}

\PHYSREV
%
%

%
\REF\NEWT{R. G. Newton, {\it Scattering Theory of Particles and Waves}, 
(McGraw-Hill Book Company, Inc., New York, 2002).}

\REF\MEHRE{R. Mehrem, J.T. Londergan and G.~E.~Walker, Phys. Rev. C {\bf 48}, 1192 (1993).}

\REF\WHITE{G. D. White, K. T. R. Davies and P. J. Siemens, Ann. 
Phys. 
   (NY) {\bf 187}, 198 (1988).}

\REF\WATSON{G. N. Watson, {\it A Treatise on the Theory of Bessel
   Functions}, (Cambridge Univ. Press, 1944).}  

\REF\TOBOC{T. Sawaguri and W. Tobocman, Journ. Math. Phys. {\bf 8}, 
2223 (1967).}

\REF\JACKSON{A. D. Jackson and L. Maximon, SIAM J. Math. Anal. {\bf 3},
   446 (1972).}

\REF\TAFFAR{R. Anni and L. Taffara, Nuovo Cimento {\bf 22A}, 12 
(1974).}

\REF\ELBAZM{E. Elbaz, J. Meyer and R. Nahabetian, Lett. Nuovo 
Cimento{\bf 10}, 418 (1974).}

\REF\ELBAZ{E. Elbaz, Riv. Nuovo Cimento {\bf 5}, 561 (1975).}

\REF\GERV{A.~Gervais and H.~Navelet, J. Math. Phys. {\bf 26}, 633, 
645(1985).}

\REF\MEHR{R. Mehrem, J.T. Londergan and M.~H.~Macfarlane, J. Phys. A {\bf 24}, 1435 (1991).}

\REF\DAVIES{K. T. R. Davies, M. R. Strayer and G. D. White, J. Phys. 
   {\bf G14}, 961 (1988).}

\REF\DAVIEST{K. T. R. Davies, J. Phys. 
   {\bf G14}, 973 (1988).}

\REF\CHEN{J. Chen and J. Su, Phys. Rev. D {\bf 69}, 076003 (2004).}

\REF\GRADST{I.S. Gradshteyn and I.M. Rhyzik: {\it Table of 
Integrals, Series and Products} (Academic Press, New York, 1965).}

\REF\BRNKSA{\rm D.M. Brink and G.R. Satchler, {\it Angular Momentum}
(Oxford University Press, London 1962).}

\REF\EDMNDS{\rm A. R. Edmonds, {\it Angular Momentum in Quantum 
Mechanics}(Princeton University Press, Princeton, 1957).}

\title{\bf THE PLANE WAVE EXPANSION, INFINITE INTEGRALS AND IDENTITIES INVOLVING SPHERICAL
BESSEL FUNCTIONS}
\author{R.\ Mehrem$^\star$} 
\address{\it Associate Lecturer \break
The Open University in the North West \break
351 Altrincham Road \break
Sharston, Manchester M22 4UN \break
United Kingdom}
\footnote{}{$\star$ Email: rami.mehrem@btopenworld.com.} 
\abstract{This paper shows that the plane wave expansion 
can be a useful tool in obtaining analytical solutions to infinite integrals over
spherical Bessel functions and the derivation of identites for these functions.  
The integrals are often used in nuclear scattering calculations, 
where an analytical result can provide an insight into the reaction mechanism.
A technique is developed whereby an integral over several special functions 
which cannot be found in any standard integral table can be broken down into 
integrals that have existing analytical solutions.}
\title{Keywords}   
Plane Wave Expansion-Infinite Integrals Over Spherical Bessel Functions-Spherical Bessel Function Identities-Special Functions-Nuclear Reactions Theory
 
\endpage
%
%
\chapter{Introduction}
The plane wave expansion, or the Rayleigh equation, is commonly used in nuclear physics; in particular it 
can be found in any standard text on scattering theory [1]. In the plane wave appproximation, 
a particle incident on a nucleus has a wavefunction approximated by 
a plane wave of the form $e^{i \vec k . \vec r}$, where $\vec k$ is the particle's wavenumber 
and $\vec r$ is its position. This approximation is often used to get an insight into the reaction mechanism 
of the process studied, without the complications arising from using the more accurate distorted waves. 

Using the plane wave approximation will result in radial integrals that involve spherical Bessel functions. 
On the other hand, there are many other applications in nuclear physics that encounter these types of integrals 
amongst which are distorted wave calculations [2] and 
nuclear response function calculations [3]. These types of integrals have in the past been calculated 
analytically (many references exist, see for example references [2] and [4-11]), or numerically 
using complex-plane methods (see for example references [12-14]). In this paper, 
an additional advantage to the plane wave expansion is presented, whereby the expansion itself is used to derive 
analytical solutions to integrals over spherical Bessel functions and identities involving them. 
The technique itself is not new [11], but the author is not aware of any prior paper that has highlighted 
its versatility.
In section 2, the plane wave expansion is introduced from which the integral representation of the spherical 
Bessel 
function is derived. Section 3 builds on that by deriving integrals involving one spherical Bessel function. 
Section 4 details integrals involving two spherical Bessel functions, amongst which is the {\it Closure Relation}
for spherical Bessel functions. Section 5 deals with integrals involving three spherical Bessel functions. 
Section 6 extends the previous results to a larger number of any special or elementary function. Here the 
Closure relation is used to simplify an integral that cannot be found in any standard integral tables [15].
This can be done, for example, by breaking an integral over 4 special functions into an integral over two 
integrals of 3 special functions each. Concluding remarks then follow.
 
\chapter{The Plane Wave Expansion}
 
The plane wave expansion, also known as the Rayleigh equation, is given by

$$e^{i\vec k \,.\,\vec r}\,= \,4\pi \,\sum_{L=0}^\infty \,\sum_{M=-L}^L\, i^L \,Y_L^M(\hat r) \, 
Y_L^{M\star}(\hat k) j_L(kr)\,\eqn\TWOPONE $$
where $Y_L^M(\hat r)$ is the spherical harmonic function for the unit vector $\hat r$, 
and $j_L(kr)$ is the spherical Bessel function for $k \ge 0$, which is assumed for the rest of the paper 
along with $k_n \ge 0$, for integer n. An alternative formula is
$$e^{i\vec k \,.\,\vec r}\,=\,\sum_{L=0}^\infty\,(i)^L\,(2L+1)\,P_L(\cos \theta)\,j_L(kr),\eqn\TWOPTWO$$
using the {\it{Vector Addition Theorem}} given by
$$P_L(\cos\theta)\,=\,{4\pi \over 2L+1}\,\sum_{M=-L}^{M=L}\,Y_L^M(\hat r)\,
Y_L^{M\star}(\hat k),\eqn\TWOPTHREE $$
where $P_L(\cos\theta)$ is the Legendre polynomial and $\theta$ is the angle between $\vec k$ and $\vec r$.

To limit the sum over $L$ in eq. $\TWOPONE$ or eq. $\TWOPTWO$ to a specific value, one needs to use 
the orthognality of the spherical harmonics
$$\int \,Y_L^M(\hat k)\,Y_{L'}^{M'\star}(\hat k)\,d\Omega_{\hat k}=\,\delta_{L,L'}\,\delta_{M,M'},\eqn\TWOPFOUR $$
where $\Omega_{\hat k}$ is the solid angle subtended by $\hat k$. Alternatively, we can use the orthognality 
of the Legendre polynomials
$$\int_{-1}^1\,P_L(\cos \theta)\,P_{L'}(\cos \theta)\,d(\cos \theta)\,=\,{2\over 2L+1}\,\delta_{L,L'}
\eqn\TWOPFIVE $$
to arrive at the integral representation of the spherical Bessel function
$$j_L(kr)\,=\,{(-i)^L\over 2}\,\int_{-1}^1\,P_L(\cos \theta)\,e^{ikr\cos \theta}\,d(\cos \theta).\eqn
\TWOPSIX $$
Setting $k\,=\,0$, one arrives at the identity
$$j_L(0)\,=\,\delta_{L,0}\eqn\TWOPSEVEN$$

\chapter{Infinite Integrals Over One Spherical Bessel Function}
Integrating eq. $\TWOPSIX$ over $r$ and interchanging the order of integration we obtain
$$\int_{-\infty}^\infty\,j_L(kr)\,dr\,=\,{(-i)^L\over 2}\,\int_{-1}^1\,P_L(\cos \theta)\,
\int_{-\infty}^\infty\,e^{ikr\cos \theta}\,dr\,d(\cos \theta).\eqn\THREEPONE$$
However,
$$\int_{-\infty}^\infty\,e^{ikr}\,dr\,=\,2\pi\,\delta(k),\eqn\THREEPTWO$$
resulting in
$$\int_{-\infty}^\infty\,j_L(kr)\,dr\,=\,{(-i)^L\over k}\,\pi\,P_L(0).\eqn\THREEPTHREE$$
Another technique is to integrate the plane wave, eq. $\TWOPONE$ over all space and using
$$\int\,d^3r\,e^{i\vec k . \vec r}\,=\,(2\pi)^3\,\delta^3(\vec k),\eqn\THREEPFOUR$$
one reaches
$$\int_0^\infty\,r^2\,j_0(kr)\,dr=\,2\pi^2\,\delta^3(\vec k),\eqn\THREEPFIVE$$
which shows the spherical symmetry of $\delta^3(\vec k)$, since there is no angular dependence on 
the left hand side.
Also, since
$$\int {e^{i\vec k\, .\,\vec r}\over r^2}\,d^3r\,=\,{2\pi^2\over k},\eqn\THREEPSIX$$
we can reach, using eqs. $\TWOPONE$ and $\TWOPFOUR$,
$$\int_0^\infty\,j_0(kr)\,dr\,=\,{\pi\over 2k}.\eqn\THREEPSEVEN$$

\chapter{Infinite Integrals Over Two Spherical Bessel Functions}
Again the plane wave expansion gives
$$\eqalign{e^{i(\vec k_1 + \vec k_2)\,.\,\vec r}\,&=\,(4\pi)^2\,\sum_{L_1,M_1}\,\sum_{L_2,M_2}\,
(i)^{L_1+L_2}\,
Y_{L_1}^{M_1}(\hat k_1)\,Y_{L_2}^{M_2\star}(\hat k_2)\cr\bspac &\times
Y_{L_1}^{M_1\star}(\hat r)\,Y_{L_2}^{M_2}(\hat r)\,j_{L_1}(k_1r)\,j_{L_2}(k_2r).}\eqn\FOURPONE$$

Integrating over all space results in fixing $L_2\,=L_1\,\equiv\,L$ and $M_2\,=M_1\,\equiv\,M$
$$\eqalign{(2\pi)^3\,\delta^3(\vec k_1+\vec k_2)\,&=\,(4\pi)^2\,\sum_{L,M}\,(-1)^L\,Y_L^M(\hat k_1)
Y_L^{M\star}(\hat k_2)\cr\bspac &\times \int_0^\infty \,r^2\,j_L(k_1r)\,j_L(k_2r)\,dr.}\eqn\FOURPTWO$$
Using
$$\delta^3(\vec k_1+\vec k_2)\,=\,{\delta(k_1\,-\,k_2)\over k_1^2}\,\delta(\Omega_{-\hat k_1}-
\Omega_{\hat k_2}),\eqn\FOURPTHREE$$
multiplying both sides of eq. $\FOURPTWO$ by $Y_L^{M\star}(\hat k_1)\,Y_L^M(\hat k_2)$, 
followed by integrating over solid angles $\hat k_1$ and $\hat k_2$ and using
$$Y_L^M(\widehat {-k})\,=\,(-1)^L\,Y_L^M(\hat k),\eqn\FOURPFOUR$$
results in
$$\int_0^\infty \,r^2\,j_L(k_1r)\,j_L(k_2r)\,dr\,=\,{\pi\over 2 k_1^2}\,\delta (k_1 \,-\, k_2),\eqn\FOURPFIVE$$
which is known as the {\it {Closure Relation}} for spherical Bessel functions. 

Another relation that can be derived using eqs. $\TWOPSIX$ and $\THREEPTWO$ is
$$\int_{-\infty}^\infty \,j_L(x)\,j_{L'}(x)\,dx\,=\,={(-i)^{L+L'}\,\pi\over 2}\,\int_{-1}^1 \,
P_L(y)\,P_{L'}(-y)\,dy.\eqn\FOURPSIX$$
Now
$$P_L(-y)\,=\,(-1)^L\,P_L(y),\eqn\FOURPSEVEN$$ 
just as 
$$j_L(-x)\,=\,(-1)^L\,j_L(x),\eqn\FOURPEIGHT$$
which reduces eq. $\FOURPSIX$ to
$$\int_{-\infty}^\infty \,j_L(x)\,j_{L'}(x)\,dx\,=\,{\pi\over 2L+1}\,\delta_{L,L'},\eqn\FOURPNINE$$
known as the {\it{Orthogonality Relation}} for the spherical Bessel functions.
Using this equation together with the plane wave expansion, eq. $\TWOPTWO$ leads to
$$P_L(x)\,=\,{(-i)^L\over \pi}\,\int_{-\infty}^\infty \,e^{ixy}\,j_L(y)\,dy,\eqn\FOURPTEN$$
which is an integral representation for the Legendre polynomial valid for $-1 < x < 1$.
Also, dividing eq. $\FOURPONE$ by $r^2$ and integrating over all space results in
$$\eqalign{{2\pi^2\over |\vec k_1 + \vec k_2 |}\,&=\,(4\pi)^2\,\sum_{L,M}\,(-1)^L\,Y_L^M(\hat k_1)
\,Y_L^{M\star}(\hat k_2)\cr\bspac &\times \int_0^\infty \,j_L(k_1r)\,j_L(k_2r)\,dr.}\eqn\FOURPELEVEN$$
Now
$${1\over |\vec k_1+\vec k_2|}\,=\,4\pi\,\sum_{L,M}\,{(-1)^L\over 2L+1}\,{(k_<)^L\over (k_>)^{L+1}}\,
Y_L^M(\hat k_1)\,Y_L^{M\star}(\hat k_2),\eqn\FOURPTWELVE$$
where $k_<$ and $k_>$ are the smaller and larger of $k_1$ and $k_2$, respectively. 
So substituting eq. $\FOURPTWELVE$ into
eq. $\FOURPELEVEN$, multiplying by $Y_L^{M\star}(\hat k_1)\,Y_L^M(\hat k_2)$ and integrating
over solid angles $\hat k_1$ and $\hat k_2$ results in
$$\int_0^\infty \,j_L(k_1r)\,j_L(k_2r)\,dr\,=\,{\pi\over 2(2L+1)}\,\,{(k_<)^L\over (k_>)^{L+1}},
\eqn\FOURPTHIRTEEN$$ 
which agrees with reference [\WATSON], equation 1, page 405.

An identity can also be derived by equating the expansion in eq. $\FOURPONE$ to the alternative expansion 
in terms of $j_L(|\vec k_1+\vec k_2|r)$ resulting in
$$\eqalign{&(i)^L\,Y_L^M(\widehat{\vec k_1+\vec k_2})\,j_L(|\vec k_1+\vec k_2|r)\,=\,\sqrt{4\pi}
\,\sum_{L_1,M_1}\,\sum_{L_2,M_2}\,(i)^{L_1+L_2}\,\sqrt{{(2L_1+1)\,(2L_2+1)\over (2L+1)}}\cr\bspac &\times \, 
<\,L_1\,L_2\,0\,0\,|\,L\,0\,>\,<\,L_1\,L_2\,M_1\,\,-M_2\,|\,L\,M\,> \,Y_{L_1}^{M_1}(\hat k_1)\,Y_{L_2}^{-M_2}
(\hat k_2)\cr\bspac &\times j_{L_1}(k_1r)\,j_{L_2}(k_2r),}\eqn\FOURPFORTEEN$$
using
$$\eqalign{\int d\Omega_{\hat r} \, Y_{l_1}^{m_1}(\hat r)\,Y_{l_2}^{m_2}(\hat r)\,Y_l^{m\star}(\hat r)\,&=\,
\sqrt{{(2l_1+1)\,(2l_2+1)\over 4\pi\,(2l+1)}}\cr\bspac &\times <\,l_1\,l_2\,0\,0\,|\,l\,0\,>\,
<\,l_1\,l_2\,m_1\,m_2\,|\,l\,m\,>,}\eqn\FOURPFIFTEEN$$
where $<\,l_1\,l_2\,m_1\,m_2\,|\,l\,m\,>$ is a Clebsch-Gordan coefficient, coupling angular momenta $l_1$ and $l_2$ 
to angular momentum $l$ such that $m_1 + m_2\,=\,m$.

If $\vec k_1$ and $\vec k_2$ point in the same direction, i.e $\hat k_1\,=\,\hat k_2\,\equiv \,\hat k$, then 
multiplying eq. $\FOURPFORTEEN$ by $Y_L^{M\star}(\hat k)$, integrating over solid angle $\hat k$ 
then summing over $M_1$ and $M_2$ using the unitarity of the Clebsch-Gordan coefficients
$$\sum_{M_1,M_2}\,<\,L_1\,L_2\,M_1\,M_2\,|\,L\,M\,>\,<\,L_1\,L_2\,M_1\,M_2\,|\,L'\,M'\,>\,=\,\delta_{L,L'}\,
\delta_{M,M'},\eqn\FOURPSIXTEEN$$
equation $\FOURPFORTEEN$ reduces to the identity
$$\eqalign{j_L [(k_1+k_2)r]\,&=\,\sum_{L_1,L_2}\,(i)^{L_1+L_2-L}\,{(2L_1+1)\,(2L_2+1)\over(2L+1)}\,
<\,L_1\,L_2\,0\,0\,|\,L\,0\,>^2 \cr\bspac &\times j_{L_1}(k_1r)\,j_{L_2}(k_2r).}\eqn\FOURPSEVENTEEN$$
Similarly,
$$\eqalign{j_L [(k_1-k_2)r]\,&=\,\sum_{L_1,L_2}\,(i)^{L_1-L_2-L}\,{(2L_1+1)\,(2L_2+1)\over(2L+1)}\,
<\,L_1\,L_2\,0\,0\,|\,L\,0\,>^2 \cr\bspac &\times j_{L_1}(k_1r)\,j_{L_2}(k_2r).}\eqn\FOURPEIGHTEEN$$
Redefining $x\equiv k_1r$ and $y\equiv k_2r$, multiplying eq. 
$\FOURPEIGHTEEN$ by 
$j_l(x)$ and integrating over x would result in an equation of the form 
$$\int_{-\infty}^\infty \, j_{L}(x-y)\,j_{l}(x)\,dx\,=\,\pi \,
\sum_{L_2} \,(i)^{l-L_2-L}\,
<\,L\, l\,0\,0\,|\,L_2\,0\,>^2\,j_{L_2}(y),\,\eqn\FOURPNINTEEN$$
using eq. $\FOURPNINE$ and the symmetry properties of 
the Clebsch-Gordan coefficients.

\chapter{Integrals Over Three Spherical Bessel Functions}
Using the same procedure as before, integrating over the expansion of $e^{i(\vec k_1 +\vec k_2 +\vec k_3).\vec r}$
and using eq. $\FOURPFIFTEEN$ amounts to
$$\eqalign{&\delta^3(\vec k_1+\vec k_2+\vec k_3)\,=\,{4\over \sqrt{\pi}}\,\sum_{L_1,M_1}\,\sum_{L_2,M_2}\,\sum_{L_3,M_3}\,
(i)^{L_1+L_2+L_3}\,\cr\bspac &\times \sqrt{{(2L_1+1)\,(2L_2+1)\over (2L_3+1)}}\,<\,L_1\,L_2\,0\,0\,|\,L_3\,0\,>\,
K(L_1,L_2,L_3,\hat k_1,\hat k_2,\hat k_3)\cr\bspac &\times
\int_0^\infty \, r^2\,j_{L_1}(k_1r)\,j_{L_2}(k_2r)\,j_{L_3}(k_3r)\,dr,}\eqn\FIVEPONE$$
where
$$\eqalign{K(L_1,L_2,L_3,\hat k_1,\hat k_2,\hat k_3)\,&\equiv \,
\sum_{M_1}\,\sum_{M_2}\,\sum_{M_3}\,
<\,L_1\,L_2\,M_1\,M_2\,|\,L_3\,M_3\,>\cr\bspac &\times
Y_{L_1}^{M_1\star}(\hat k_1)\,Y_{L_2}^{M_2\star}(\hat k_2)\,
Y_{L_3}^{M_3}(\hat k_3).}\eqn\FIVEPTWO$$
Multiplying both sides of eq. $\FIVEPONE$ by 
$K^\star (L^\prime_1,L^\prime_2,L^\prime_3,\hat k_1,\hat k_2,\hat k_3)$ and using
the orthognality of the spherical harmonics
$$\eqalign{&\int d\Omega_{\hat k_1} d\Omega_{\hat k_2} d\Omega_{\hat k_3} K^\star (L'_1,L'_2,L'_3,\hat k_1,\hat k_2,\hat k_3)\,
K(L_1,L_2,L_3,\hat k_1,\hat k_2,\hat k_3) \,\cr\bspac &=\,(2L_3+1)\,\delta_{L'_1,L_1}\,\delta_{L'_2,L_2}\,
\delta_{L'_3,L_3},}\,\eqn\FIVEPTHREE$$
results in
$$\eqalign{&<\,L_1\, L_2\,0\,0\,|\,L\,0\,>\,\int_0^\infty \, r^2\,j_{L_1}(k_1r)\,j_{L_2}(k_2r)\,j_{L_3}(k_3r)\,dr
\cr\bspac &=\,{\sqrt {\pi}\over 4}\,{(-i)^{L_1+L_2+L_3}\over 
\sqrt{(2L_1+1)(2L_2+2)(2L_3+1)}}\cr\bspac &\times
\int d \Omega_{\hat k_1} d \Omega_{\hat k_2}
 d \Omega_{\hat k_3}\,   
K^\star (L_1,L_2,L_3,\hat k_1,\hat k_2,\hat k_3)\,\delta^3 (\vec k_1 + \vec k_2 + \vec k_3).}\eqn\FIVEPFOUR$$
Now
$$\delta^3 (\vec k_1+\vec k_2 + \vec k_3)\,=\,{\delta (k_3 -|\vec k_1+\vec k_2|)\over k_3^2}\,\,
\delta (\Omega_{-\hat k_3}
-\Omega_{\widehat {\vec k_1 +\vec k_2}}).\eqn\FIVEPFIVE$$
Using
$$\delta (k_3 -|\vec k_1 +\vec k_2|)\,=\,{k_3 \over k_1\,k_2}\,
\,\delta (\cos \theta - {k_1^2+k_2^2-k_3^2\over 2 k_1\,k_2}),\eqn\FIVEPSIX$$
where $\theta$ is the angle between $\hat k_1$ and $\hat k_2$, and the completeness relation for 
the Legendre polynomials
$$\delta (x-y)\,=\,\sum_L\,{2L+1\over 2}\,P_L(x)\,P_L(y),\eqn\FIVEPSEVEN$$
gives
$$\delta^3 (\vec k_1 +\vec k_2+\vec k_3)\,=\,{\beta(\Delta)\over 2 k_1 k_2 k_3}\,\delta (\Omega_{-\hat k_3}
-\Omega_{\widehat {\vec k_1 +\vec k_2}})\sum_L \,(2L+1)\,P_L(\Delta)\,P_L(\cos \theta),\eqn\FIVEPEIGHT$$
where 
$$\Delta \,=\,{k_1^2+k_2^2-k_3^2\over 2k_1 k_2},\eqn\FIVEPNINE$$ 
and the factor $\beta (\Delta)$ is given by [11]
$$\eqalign {\beta (\Delta)\,&=\,{1\over 2},\,\,\,\,\,\, \,\,\,\, \Delta \,=\,\pm 1 \cr &= 1,\,\,\,\,\,\,\,\,\,\, 
-1< \Delta< 1 \cr
&=0,\,\,\,\,\,\,\,\,\,\,{\rm otherwise}.}\eqn\FIVETEN$$
Equation $\FIVEPFOUR$ then becomes
$$\eqalign{&<\,L_1\, L_2\,0\,0\,|\,L\,0\,>\,\int_0^\infty \, r^2\,j_{L_1}(k_1r)\,j_{L_2}(k_2r)\,j_{L_3}(k_3r)\,dr
\cr\bspac &=\,{\sqrt {\pi}\,\beta (\Delta)\over 8 k_1 k_2 k_3}\,\,{(-i)^{L_1+L_2+L_3}\over 
\sqrt{(2L_1+1)(2L_2+2)(2L_3+1)}}\cr\bspac &\times\int d \Omega_{\hat k_1} d \Omega_{\hat k_2} \,   
K^\star (L_1,L_2,L_3,\hat k_1,\hat k_2,\widehat {-(\vec k_1+\vec k_2)})
\sum_L \,(2L+1)\,P_L(\Delta)\,P_L(\cos(\theta)).}\eqn\FIVEPELEVEN$$
The integral in eq. $\FIVEPELEVEN$ can be easily evaluated using eq. $\TWOPTHREE$, eq. $\FOURPFOUR$
and the {\it Solid-Harmonic Addition Theorem} for spherical harmonics [16]
$$\eqalign{ &Y_{L_3}^{M_3}(\widehat {\vec k_1 +\vec k_2})\,=\,\left({k_1\over k_3}\right)^{L_3}\,
\sum_{L=0}^{L_3}\,{\sqrt{4\pi\over 2L+1}}\,{2L_3+1 \choose 2L}^{1/2}\cr\bspac &\times
\left ({k_2\over k_1}\right )^L\,
\sum_M \,<(L_3-L)\,L\,(M_3-M)\,M|L_3\,M_3>\,Y_{L_3-L}^{M_3-M}(\hat k_1)\,Y_L^M(\hat k_2),}\eqn\FIVEPTWELVE$$
where the binomial coefficient is defined by
$${n \choose m}\,\equiv \,{n!\over (n-m)! m!}.\eqn\FIVEPTHIRTEEN$$
The result of the integration is then the formula for the integral over three spherical Bessel functions
$$\eqalign{& <\,L_1\,L_2\,0\,0|L_3\,0\,>\,\int_0^\infty \, r^2\,j_{L_1}(k_1r)\,j_{L_2}(k_2r)\,j_{L_3}(k_3r)\,dr \cr\bspac
&=\,{\pi \beta (\Delta)\over 4 k_1 k_2 k_3}\,(i)^{L_1+L_2+L_3}\,(2L_3+1)\,\left ({k_1\over k_3}\right )^{L_3}\,
\sum_{L=0}^{L_3}\,{2L_3 \choose 2L}^{1/2}\,\left ({k_2\over k_1 }\right )^L\cr\bspac &\times \,\sum_l\,
<\,L_1\,(L_3-L)\,0\,0|l\,0\,>\,<\,L_2\,L\,0\,0|l\,0\,>\,\sixj {L_1}{L_2}{L_3}{L}{L_3-L}{l}\,P_l(\Delta),}
\eqn\FIVEPFOURTEEN$$
where $\sixj {j_1}{j_2}{j_3}{j_4}{j_5}{j_6}$ is a 6-j symbol defined in any standard angular momentum text [17]. 
An interesting special case of the above equation is when $L_3\,=\,0$, which imposes 
$L_1\,=\,L_2\,\equiv \,\lambda$ leading to
$$\int_0^\infty r^2\,j_\lambda(k_1r)\,j_\lambda(k_2r)\,j_0(k_3r)\,dr\,=\,{\pi \, \beta(\Delta)\over 4 k_1 k_2 k_3}\,
P_\lambda \left ({k_1^2 +k_2^2 -k_3^2 \over 2k_1 k_2}\right ),\eqn\FIVEPFIFTEEN$$
and the inverse equation 
$$j_\lambda (k_1r)\,j_\lambda (k_2r)\,=\,{1\over 2 k_1 k_2}\,\int_{|k_2-k_1|}^{k_1+k_2}\,k_3\,j_0(k_3r)\,
P_\lambda \left ({k_1^2 +k_2^2 - k_3^2\over 2 k_1 k_2}\right ) dk_3.\eqn\FIVEPSIXTEEN$$

\chapter {Generalisation}
It is possible to extend the results of the last section to finding analytical solutions to 
infinite integrals over 
several special or elementary functions (provided the integral exists) if the integral over a fewer number of these functions 
combined with a spherical Bessel function is known. 
One can show, using the {\it Closure Relation} for spherical Bessel functions eq. $\FOURPFIVE$, that
$$\eqalign{&\int_0^\infty r^2 \,\chi_{L_1}(k_1,r)\,\chi_{L_2}(k_2,r)\,\chi_{L_3}(k_3,r)\,\chi_{L_4}(k_4,r)dr\,=\,
{2\over \pi}\int_0^\infty k^2 dk \cr\bspac &\times\left (\int_0^\infty r^2\,\chi_{L_1}(k_1,r)\,\chi_{L_2}(k_2,r)\,j_L(kr)dr
\right )\,\left (\int_0^\infty r'^2\,\chi_{L_3}(k_3,r')\,\chi_{L_4}(k_4,r')\,j_L(kr') dr' \right ),}
\eqn\SIXPONE$$
where $\chi_L(k,r)$ is a special function of order $L$ that, in general, depends on $k$ and $r$. 
So, in effect, the integral over four special functions has been broken up into two integrals over 3 special functions. 
The final integral over $k$ may be easier to handle than the initial integral, or in the least is 
an alternative approach to solving the integral analytically.
The following example illustrates the advantage:

Example- Evaluate the following integral over 4 spherical Bessel functions analytically
$$I\,\equiv\,\int_0^\infty r^2 j_1(k r)\,j_1(k r)\,j_2(k r)\,j_2(k r)dr.\eqn\SIXPTWO$$ 

Solution- The integral can be rewritten as
$$I\,=\,{2\over \pi}\,\int_0^\infty q^2 dq\,\left (\int_0^\infty r^2 j_1(kr)\,j_1(kr)\,j_0(qr)dr
\right ) \left (\int_0^\infty r'^2 j_2(kr')\,j_2(kr')\,j_0(qr') dr'\right ).\eqn\SIXPTHREE$$
Using eq. $\FIVEPFIFTEEN$ this reduces to
$$I\,=\,{\pi\over 8k^4}\,\int_0^{2k}\,
P_1\left ({2k^2-q^2\over 2 k^2}\right )\,P_2\left ({2k^2-q^2\over 2 k^2}\right )dq.
\eqn\SIXPFOUR$$
This can be easily evaluated using $P_1(x)=x$ and $P_2(x)={1\over 2}\,(3x^2-1)$ to give
$$I\,=\,{23\pi \over 420 k^3}.\eqn\SIXPFIVE$$
\endpage

\chapter{Conclusions}
The plane wave expansion was utilised to derive integrals over one, two, and three spherical Bessel functions. 
The generalisation to a larger number of special functions is exhibited through the use of 
the {\it Closure Relation} for spherical Bessel functions, which is also derived here using 
the plane wave expansion. In addition, several identities involving spherical Bessel functions were also derived 
using the same techniques.
\chapter{References}
[1] R. G. Newton, Scattering Theory of Particles and Waves, 
McGraw-Hill Book Company, Inc., New York, 2002.

[2] R. Mehrem, J.T. Londergan and G.~E.~Walker, Phys. Rev. C48 (1993) 1192.

[3] G. D. White, K. T. R. Davies and P. J. Siemens, Ann. Phys. 
   (NY) 187 (1988) 198.

[4] G. N. Watson, A Treatise on the Theory of Bessel
   Functions, Cambridge Univ. Press, 1944.  

[5] T. Sawaguri and W. Tobocman, Journ. Math. Phys. 8 (1967) 2223.

[6] A. D. Jackson and L. Maximon, SIAM J. Math. Anal. 3 (1972) 446.

[7] R. Anni and L. Taffara, Nuovo Cimento 22A (1974) 12.

[8] E. Elbaz, J. Meyer and R. Nahabetian, Lett. Nuovo Cimento 10 (1974) 418.

[9] E. Elbaz, Riv. Nuovo Cimento 5 (1975) 561.

[10] A.~Gervais and H.~Navelet, J. Math. Phys. 26 (1985) 633, 645.

[11] R. Mehrem, J.T. Londergan and M.~H.~Macfarlane, J. Phys. A 24 (1991) 1435.

[12] K. T. R. Davies, M. R. Strayer and G. D. White, J. Phys. G14 (1988) 961.

[13] K. T. R. Davies, J. Phys. G14 (1988) 973.

[14] J. Chen and J. Su, Phys. Rev. D 69 (2004) 076003.

[15] I.S. Gradshteyn and I.M. Rhyzik: Table of Integrals, 
Series and Products, Academic Press, New York, 1965.

[16] D.M. Brink and G.R. Satchler, Angular Momentum,
Oxford University Press, London 1962.

[17] A. R. Edmonds, Angular Momentum in Quantum Mechanics,
Princeton University Press, Princeton, 1957.

\endpage
\bye